\documentclass[10pt,letterpaper]{article}
\usepackage{spconf,amsmath,amssymb,graphicx,booktabs,cite}
\usepackage[T1]{fontenc}
\usepackage{tikz,xurl,flushend}
\usetikzlibrary{arrows.meta,positioning,calc}
\usepackage[hidelinks]{hyperref}
\newcommand{\mat}[1]{\mathbf{#1}}
\newcommand{\vect}[1]{\boldsymbol{#1}}
\newcommand{\up}{\ensuremath{\uparrow}}
\newcommand{\dn}{\ensuremath{\downarrow}}
\title{Signal-Independent and Signal-Dependent\\Neural Ambisonic Matrix Encoding for Arbitrary Arrays\\with Variable Microphone Counts}
\name{Shichao Hu$^{1,2}$, Zhiheng Jin$^{1,2}$, Chunyang Xu$^{1,2}$, Mengyao Zhu$^{1,2,\dagger}$%
\thanks{$^{\dagger}$Corresponding author: \texttt{zhu.mengyao@suda.edu.cn}.}}
\address{$^1$Center of Auditory and Language Intelligence, Soochow University, Suzhou, China\\
$^2$School of Future Science and Engineering, Soochow University, Suzhou, China}
\begin{document}
\ninept
\renewcommand{\footnotesize}{\fontsize{9}{10.5}\selectfont}
\maketitle
\raggedbottom

\begin{abstract}
Recent neural Ambisonic encoders accommodate diverse array geometries, yet many existing neural encoders require a fixed microphone count because the number of microphone channels is embedded in the network architecture. This requirement limits deployment across devices with different microphone configurations and adaptation to changes in available channels. To address this limitation, we investigate Transformer-based matrix encoding for arbitrary microphone arrays with variable microphone counts. This is achieved through shared microphone-wise processing and masked self-attention that models inter-microphone relationships across variable-size arrays. Within this framework, we consider signal-independent (SI) encoding, which predicts encoding matrices from array transfer functions, and introduce a signal-dependent (SD) extension that additionally incorporates the observed microphone signals. Both models are trained on simulated scenes using LibriSpeech sources and extensively evaluated under changes in source type, unseen microphone counts, and increased source counts beyond those used during training. Both SI and SD outperform conventional least-squares (LS) encoding in aggregate reconstruction performance across the evaluated conditions. SD consistently achieves stronger overall performance than SI. These results demonstrate that the proposed framework enables array-agnostic Ambisonic encoding while retaining generalization across microphone counts and acoustic source conditions.
\end{abstract}
\begin{keywords}
Ambisonics, array transfer functions, variable microphone count, signal-dependent encoding, generalization
\end{keywords}

\section{Introduction}
\label{sec:intro}
Spatial audio systems benefit from a common sound-field representation across devices. In wearable listening, robot audition, mobile recording, and augmented reality, physical design and hardware constraints lead to different microphone layouts and counts; sensor failures may further change available inputs. Ambisonics represents sound fields through spherical-harmonic (SH) coefficients~\cite{zotter,rafaelybook}, decoupling reproduction from capture geometry. Beyond rendering, first-order Ambisonics is also emerging as a spatial audio format for localization and spatial reasoning in language models~\cite{spatialomni}, further motivating a device-independent encoding front end that supports diverse microphone configurations.

Classical signal-independent encoders match modeled or measured array transfer functions (ATFs) to SH responses using least-squares (LS) optimization~\cite{politis,gayer,bastine}. The resulting filters accommodate diverse geometries but require recomputation when the array changes. Regularization balances response accuracy against noise amplification, while limited microphone counts, unfavorable geometries, and spatial aliasing constrain reconstruction~\cite{gayer,aliasing}. These filters remain fixed during inference and do not adapt to source activity. In contrast, parametric encoding uses a multi-source sound-field model to separate source and directional ambient components before Ambisonics encoding~\cite{parametric}. This signal-dependent approach can improve bandwidth and spatial resolution, but relies on scene-model assumptions and accurate spatial parameter estimation~\cite{parametric}.

These limitations motivate neural encoders that learn signal statistics to predict Ambisonic signals or encoding weights~\cite{neural2024,qiao,residual}. Furthermore, to address unseen array layouts, Gen-A conditions encoding on microphone coordinates~\cite{gena}, whereas Beyond Omnidirectional uses directional ATFs and cross-attention to model directivity and device scattering~\cite{beyond}; both retain a fixed microphone count. Generative approaches include DiffM2A, which combines modal projection with conditional diffusion~\cite{diffm2a}, and ADEPS, which supports different counts by coupling an array-independent Ambisonic prior with an array-dependent observation operator~\cite{adeps}. However, multi-step reverse diffusion requires repeated network evaluations, increasing inference cost and posing challenges for real-time encoding~\cite{diffm2a,adeps}. Flow-HOA instead generates time-invariant FIR encoders offline~\cite{flowhoa}. These advances address different aspects of geometry adaptation, count flexibility, and source-domain generalization.

In this study, we focus on neural matrix encoding that supports variable microphone counts. We use shared microphone-wise processing, masked self-attention~\cite{settransformer}, and a shared coefficient head to accommodate different microphone counts without modifying the network architecture. Building on our ATF-conditioned signal-independent (SI) encoder~\cite{si_previous}, we introduce a signal-dependent (SD) extension incorporating microphone signals. Unlike iterative diffusion inference, both predictors use a single feed-forward pass and are fully causal in the STFT-frame domain, supporting streaming-oriented processing. To assess the generalization ability beyond the training conditions, we train on simulated LibriSpeech scenes and evaluate unseen microphone counts, denser source mixtures, and non-speech sources without retraining. Experiments show that both SI and SD retain improvements over Static LS under the evaluated unseen conditions. Incorporating microphone signals in SD further improves overall FOA reconstruction over SI.

\section{Methods}
\label{sec:method}
\subsection{Baseline Method}
Let $\vect{x}_{ft}\in\mathbb{C}^{P}$ denote the short-time Fourier transform (STFT) of a $P$-microphone mixture, and $\vect{b}_{ft}\in\mathbb{C}^{4}$ its target first-order Ambisonic (FOA) coefficients. The desired coefficients describe the total scene, not separated sources. We use real ACN/SN3D channels $[W,Y,Z,X]$. For a unit direction $\vect{u}_d$, the FOA steering vector is
\begin{equation}
 \vect{y}_d=[1,u_{d,y},u_{d,z},u_{d,x}]^{\mathsf T}.
\end{equation}
Collecting $D$ directions gives $\mat{Y}\in\mathbb{R}^{4\times D}$ and the array-specific ATF matrix $\mat{H}_f\in\mathbb{C}^{P\times D}$. The regularized least-squares (LS) encoder is
\begin{equation}
 \mat{E}^{\mathrm{LS}}_f=\mat{Y}\mat{H}_f^{\mathsf H}
 (\mat{H}_f\mat{H}_f^{\mathsf H}+\lambda\mat{I}_P)^{-1}.
 \label{eq:ls}
\end{equation}
Here $\lambda$ is the Tikhonov regularization parameter. The encoder provides $\widehat{\vect{b}}^{\mathrm{LS}}_{ft}=\mat{E}^{\mathrm{LS}}_f\vect{x}_{ft}$. In this study, the directional grid consists of $D=96$ approximately uniformly distributed Fibonacci samples. 

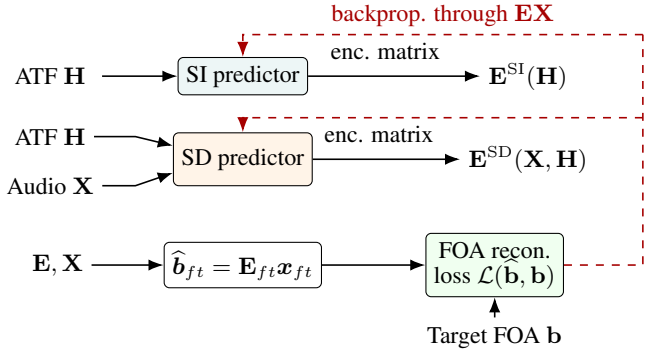
\begin{figure}[ht]
\centering
\begin{tikzpicture}[x=.98cm,y=1cm,>=Latex,font=\fontsize{9}{10}\selectfont,
 block/.style={draw,rounded corners=2pt,align=center,inner sep=3pt},
 flow/.style={->,line width=.6pt},grad/.style={->,dashed,draw=red!65!black,line width=.6pt}]
\node at (.5,1.6) {ATF $\mat H$};
\node[block,fill=teal!7] (si) at (3.1,1.6) {SI predictor};
\node (esi) at (6.95,1.6) {$\mat E^{\rm SI}(\mat H)$};
\draw[flow] (1.2,1.6)--(si);
\draw[flow] (si)--node[pos=.45,above=3pt]{enc. matrix}(esi);
\node (h) at (.5,.8) {ATF $\mat H$};
\node (x) at (.5,.15) {Audio $\mat X$};
\node[block,fill=orange!9,minimum height=.70cm] (sd) at (3.1,.5) {SD predictor};
\node (esd) at (6.95,.5) {$\mat E^{\rm SD}(\mat X,\mat H)$};
\draw[flow] (h.east)--(1.65,.8)--($(sd.west)+(0,.18)$);
\draw[flow] (x.east)--(1.65,.15)--($(sd.west)-(0,.18)$);
\draw[flow] (sd)--node[pos=.45,above=3pt]{enc. matrix}(esd);
\node (inputs) at (.6,-.9) {$\mat E,\mat X$};
\node[block] (apply) at (3.1,-.9) {$\widehat{\vect b}_{ft}=\mat E_{ft}\vect x_{ft}$};
\node[block,fill=green!6] (loss) at (6.5,-.9) {FOA recon.\\loss $\mathcal L(\widehat{\mat b},\mat b)$};
\draw[flow] (inputs)--(apply);
\draw[flow] (apply)--(loss);
\node (target) at (6.5,-1.85) {Target FOA $\mat b$};
\draw[flow] (target)--(loss);
\draw[dashed,draw=red!65!black,line width=.6pt] (loss.east)--(8.5,-.9)--(8.5,2.15);
\draw[grad] (8.5,2.15)--node[above,text=red!65!black]{backprop. through $\mat E\mat X$}(3.1,2.15)--(si.north);
\draw[grad] (8.5,1.05)--(3.1,1.05)--(sd.north);
\end{tikzpicture}
\caption{SI versus SD matrix prediction and shared training. $\mat E$ includes the analytical LS term. Solid arrows show forward computation; dashed arrows indicate loss gradients to the selected predictor. SI uses audio for reconstruction supervision, not as predictor input. The loss supervises FOA, not separated sources.}
\label{fig:si_sd}
\end{figure}

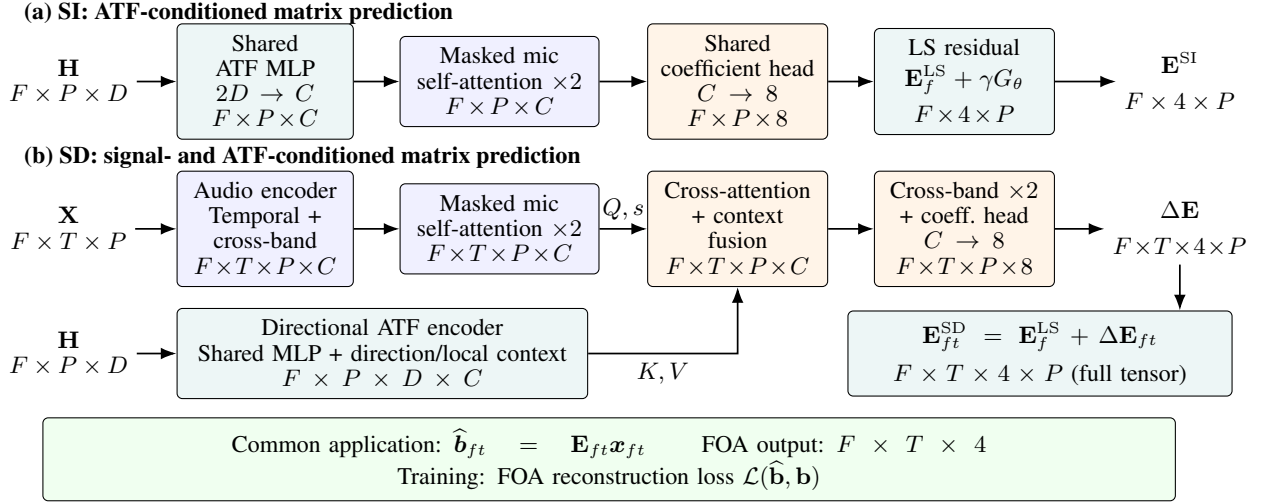
\begin{figure*}[t]
\centering
\begin{tikzpicture}[x=1cm,y=1cm,>=Latex,font=\fontsize{9}{10}\selectfont,
 box/.style={draw,rounded corners=2pt,align=center,inner sep=4pt,minimum height=.85cm},
 atf/.style={box,fill=teal!7},aud/.style={box,fill=blue!6},
 dec/.style={box,fill=orange!9},arr/.style={->,line width=.65pt}]
 \node[anchor=west,font=\bfseries] at (-.2,4.45) {(a) SI: ATF-conditioned matrix prediction};
 \node[align=center] (sh) at (.5,3.55) {$\mat H$\\$F\times P\times D$};
 \node[atf,text width=2.05cm] (se) at (3.1,3.55)
 {\mbox{Shared}\\\mbox{ATF MLP}\\$2D\to C$\\$F\!\times\!P\!\times\!C$};
 \node[aud,text width=2.35cm] (sa) at (6.2,3.55)
 {\mbox{Masked mic}\\\mbox{self-attention $\times2$}\\$F\!\times\!P\!\times\!C$};
 \node[dec,text width=2.1cm] (sc) at (9.35,3.55)
 {\mbox{Shared}\\\mbox{coefficient head}\\$C\to8$\\$F\!\times\!P\!\times\!8$};
 \node[atf,text width=2.1cm] (sr) at (12.35,3.55)
 {\mbox{LS residual}\\[2pt]$\mat E_f^{\rm LS}+\gamma G_\theta$\\[4pt]$F\!\times\!4\!\times\!P$};
 \node[align=center] (so) at (15.2,3.55)
 {$\mat E^{\rm SI}$\\[4pt]$F\times4\times P$};
 \draw[arr] (sh)--(se);
 \draw[arr] (se)--(sa);
 \draw[arr] (sa)--(sc);
 \draw[arr] (sc)--(sr);
 \draw[arr] (sr)--(so);
 \node[anchor=west,font=\bfseries] at (-.2,2.55) {(b) SD: signal- and ATF-conditioned matrix prediction};
 \node[align=center] (sx) at (.5,1.6) {$\mat X$\\$F\times T\times P$};
 \node[aud,text width=2.05cm] (ae) at (3.1,1.6)
 {\mbox{Audio encoder}\\\mbox{Temporal +}\\\mbox{cross-band}\\$F\!\times\!T\!\times\!P\!\times\!C$};
 \node[aud,text width=2.35cm] (ma) at (6.2,1.6)
 {\mbox{Masked mic}\\\mbox{self-attention $\times2$}\\$F\!\times\!T\!\times\!P\!\times\!C$};
 \node[dec,text width=2.1cm] (ca) at (9.35,1.6)
 {\mbox{Cross-attention}\\\mbox{+ context}\\\mbox{fusion}\\$F\!\times\!T\!\times\!P\!\times\!C$};
 \node[dec,text width=2.1cm] (cd) at (12.35,1.6)
 {\mbox{Cross-band $\times2$}\\\mbox{+ coeff.\ head}\\$C\to8$\\$F\!\times\!T\!\times\!P\!\times\!8$};
 \node[align=center] (do) at (15.2,1.6)
 {$\Delta\mat E$\\[4pt]$F\!\times\!T\!\times\!4\!\times\!P$};
 \draw[arr] (sx)--(ae);
 \draw[arr] (ae)--(ma);
 \draw[arr] (ma)--node[above] {$Q,s$}(ca);
 \draw[arr] (ca)--(cd);
 \draw[arr] (cd)--(do);
 \node[align=center] (dh) at (.5,-.05)
 {$\mat H$\\$F\times P\times D$};
 \node[atf,text width=5.15cm] (de) at (4.65,-.05)
 {Directional ATF encoder\\Shared MLP + direction/local context\\$F\times P\times D\times C$};
 \draw[arr] (dh)--(de);
 \draw[arr] (de.east)--node[below] {$K,V$}(9.35,-.05)--(ca.south);
 \node[atf,text width=4.8cm] (dm) at (13.35,-.05)
 {$\mat E^{\rm SD}_{ft}=\mat E^{\rm LS}_f+\Delta\mat E_{ft}$\\[5pt]$F\times T\times4\times P$ (full tensor)};
 \draw[arr] (do.south)--(15.2,.65)--(dm.north -| do.south);
 \node[box,fill=green!6,text width=14.7cm] at (7.65,-1.45)
 {Common application: $\widehat{\vect b}_{ft}=\mat E_{ft}\vect x_{ft}$
 \qquad FOA output: $F\times T\times4$\\
 Training: FOA reconstruction loss $\mathcal L(\widehat{\mat b},\mat b)$};
\end{tikzpicture}
\caption{SI and SD architecture overview. Microphone count $P$ is variable, $D=96$, $F=129$, and hidden size $C=64$; batch dimensions are omitted. SI summarizes each microphone's directional ATF into one token; SD retains $D$ directional tokens as keys/values and uses audio context $s$ to form queries. Both predict four complex coefficients per microphone and use an analytical LS residual path.}
\label{fig:framework}
\end{figure*}

\subsection{Signal-independent neural encoding}
\label{sec:si}
The SI predictor we proposed~\cite{si_previous} comprises a shared ATF encoder, a masked Transformer, and a microphone-wise coefficient decoder (Fig.~\ref{fig:framework}(a)). For each microphone and frequency, the real and imaginary parts of its $D$ directional responses are concatenated into a $2D$-dimensional vector. A shared MLP maps this vector through $2D\!\to\!64\!\to\!64$ linear layers, with LayerNorm after the first layer and ReLU after both layers. Its weights are shared across microphones and frequencies.

At each frequency, two pre-normalized Transformer layers process the $P$ microphone tokens using four-head self-attention and 128-dimensional feed-forward blocks with ReLU activation and no dropout. Padding masks exclude absent microphones from attention. A shared $64\!\to\!64\!\to\!8$ MLP with an intermediate ReLU decodes each contextualized token into the real and imaginary parts of four FOA residual coefficients. These form $G_\theta(\mat H_f)\in\mathbb C^{4\times P}$, giving
\begin{equation}
 \widehat{\vect{b}}^{\rm SI}_{ft}
 =\left[\mat{E}^{\rm LS}_f+\gamma G_\theta(\mat{H}_f)\right]\vect{x}_{ft}.
 \label{eq:si}
\end{equation}
Here $\gamma$ scales the residual. The final decoder layer is zero-initialized when training from scratch, recovering LS initially. Microphone signals are used only to apply the predicted matrix and compute the FOA reconstruction loss, not as predictor inputs.

\subsection{Signal-dependent neural encoding}
\label{sec:sd}
Figure~\ref{fig:framework}(b) summarizes our proposed SD. It predicts a residual correction to the LS encoding matrix from the observed microphone spectra and the ATFs:
\begin{equation}
 \widehat{\vect{b}}^{\rm SD}_{ft}
 =\mat{E}^{\rm LS}_f\vect{x}_{ft}
 +\Delta\mat{E}_{\theta,ft}(\mat X,\mat H)\vect{x}_{ft}.
 \label{eq:sd}
\end{equation}
 This follows the residual-learning principle in~\cite{residual}. Unlike SI, the encoding matrix adapts to the observed scene through $\mat X$, as in adaptive matrix encoders~\cite{beyond}. The audio branch maps the real and imaginary components of each microphone's STFT to 64-dimensional features with two shared causal temporal convolutions and a cross-band block~\cite{spatialnet} to model inter-frequency patterns. The temporal convolutions use kernel size 3, dilations 1 and 2, left padding, and GELU activations. Each cross-band block uses two grouped frequency convolutions (kernel size 5, 8 groups) surrounding a full-band mapping with 8 bottleneck channels and channel-wise $129\times129$ frequency weights. Two pre-normalized masked Transformer layers, each with four attention heads, a 128-dimensional feed-forward block, and GELU activation, model inter-microphone relationships at each $(f,t)$, producing audio context $s$.

The ATF branch uses a shared $2\to64\to64$ MLP, learned direction embeddings, and mean aggregation over the eight nearest grid directions, including the current direction, to form key and value matrices, each of size $D\times64$, for each microphone and frequency. Four-head cross-attention uses $s$ to query these directional features. Its output is concatenated with $s$ and projected from 128 to 64 dimensions to fuse audio and ATF features. Adapted from Beyond Omnidirectional~\cite{beyond}, this mechanism retains $P$ and shares processing across microphones to support variable counts.

Two further cross-band blocks refine the fused features through the local and full-band frequency mixing. A shared $64\!\to\!64\!\to\!8$ head predicts four complex residual coefficients per microphone. Its final layer is zero-initialized, so the model initially recovers LS. The analytical LS path joins only at the output, not as a learned feature.

\subsection{Variable-count processing and masking}
Both models use shared microphone-wise processing and masked self-attention, with hidden size 64 and four attention heads. Microphone count determines the token-sequence length rather than the feature width. Padding masks exclude invalid microphones from attention, and their signals and output coefficients are zeroed. Together with shared coefficient heads, this design supports variable microphone counts without changing network parameters, while the ATFs provide array-specific acoustic information.


\section{Experimental Setup}
\label{sec:protocol}
\subsection{Compared methods}
We compare analytical LS encoding with the ATF-only SI encoder and the SD extension. All methods use the same array ATFs, microphone recordings, and FOA targets. SI and SD are trained with LibriSpeech sources and evaluated without target-domain fine-tuning.

\subsection{Data and generalization tests}
Training scenes contain one or two LibriSpeech sources~\cite{libri} and arrays with microphone counts $P\in\{4,6,8\}$, using free-field or rigid-sphere responses~\cite{duda}. Across training, validation, and testing, array radii range from 0.04 to 0.10 m, with a minimum microphone spacing of 0.01 m. Training source-level differences span $\pm15$ dB. We use 30 training and 10 validation shoebox rooms, with dimensions ranging from $5\times5\times4$ to $12\times12\times8$ m and reverberation times of 0.10--0.50 s. Training and validation use disjoint sets of 201 and 25 speakers, respectively. Each array type and microphone count uses 100 training and 20 validation geometries.

To evaluate the trained models' generalization, we independently generate test microphone signals from LibriSpeech, ESC-50~\cite{esc}, and Gaussian white noise under comparable simulation conditions. Table~\ref{tab:generalization} evaluates microphone- and source-count generalization separately on LibriSpeech. Microphone-count tests compare seen counts $P=4,6$ with unseen counts $P=5,7$, using $K=1,2$. The unseen counts are obtained by removing one microphone from 6- and 8-microphone arrays while keeping sources, rooms, and targets unchanged. Source-count tests compare $K=1,2$ with unseen counts $K=5,7$, using $P=4,6$. Microphone-count tests use 252 scenes per group; source-count tests use 252 and 336 scenes for seen and unseen counts, respectively. Each group is averaged equally over its tested count combinations.

Table~\ref{tab:domain} evaluates cross-source-domain generalization by comparing LibriSpeech with unseen ESC-50 and white-noise sources. All domains use $P=4,6$ and $K=5,7$, with 336 LibriSpeech scenes and 399 scenes each for ESC-50 and white noise. Results are averaged equally over the four $(P,K)$ combinations. Thus, domain transfer is evaluated with source counts beyond training. In Table~\ref{tab:domain}, ``seen'' denotes the source domain used during training.

\subsection{Training settings}
Audio is sampled at 16 kHz and processed in 2.5-s excerpts. STFT analysis uses a 256-point FFT, a 128-sample periodic Hann window, and hop 64. We set $\lambda=10^{-4}$ for LS and $\gamma=0.1$ for SI. Each training epoch uses 5,040 examples generated with epoch-dependent sampling, with a fixed validation set of 504 examples.

Both models minimize the same multi-resolution complex STFT MAE:
\begin{equation}
 \mathcal L=\frac{1}{2}\sum_{r\in\{256,512\}}
 \operatorname{mean}\left|\mathcal S_r(\widehat{\mat b})-
 \mathcal S_r(\mat b)\right|,
\end{equation}
where $\mathcal S_r$ uses a normalized periodic-Hann STFT with hop $r/4$. Preliminary experiments favored this objective for spectral reconstruction, coherence, and SI-SDR; adding magnitude-based terms reduced magnitude error but degraded performance on other metrics.

\subsection{Evaluation metrics}
We evaluate FOA reconstruction using waveform SI-SDR~\cite{sdr}, complex spectral mean squared error (MSE), and magnitude-squared coherence (Coh.). SI-SDR measures waveform fidelity in dB, while $\mathrm{MSE}=\operatorname{mean}_{c,f,t}|\widehat B_{cft}-B_{cft}|^2$ measures complex-spectrum error over FOA channels and time--frequency bins. Coh.\ measures agreement between corresponding predicted and target channels. 

For binaural evaluation, predicted and target FOA share an ACN/SN3D decoder fitted by regularized complex LS to MIT KEMAR HRIRs~\cite{kemar}, resampled to 16 kHz. This differs from MagLS rendering~\cite{magls}. Bin.\ $\mathrm{ILD}_{\rm Err}$ and Bin.\ $\mathrm{IC}_{\rm Err}$ denote mean absolute prediction--reference errors in interaural level difference (dB) and interaural coherence, respectively. Both errors use decoded target FOA as the reference, not measured binaural ground truth. Scores are computed per scene before group averaging.

\section{Results and Discussion}
\label{sec:results}

\begin{table}[t]
\centering
\caption{LibriSpeech results averaged equally over the count combinations in each group. Microphone-count tests use seen source counts $K=1,2$; source-count tests use seen microphone counts $P=4,6$. SI-SDR and ILD error are in dB.}
\label{tab:generalization}
\setlength{\tabcolsep}{2pt}
\renewcommand{\arraystretch}{1.0}
\begin{tabular*}{\dimexpr\columnwidth-8pt\relax}{@{\extracolsep{\fill}}lrrr@{\hspace{0pt}}rr@{}}
\toprule
Method & SI-SDR\up & MSE\dn & Coh.\up & \shortstack{Bin.\\$\mathrm{ILD}_{\rm Err}$\dn} & \shortstack{Bin.\\$\mathrm{IC}_{\rm Err}$\dn}\\
\midrule
\multicolumn{6}{l}{Mic.~count: $P_{\rm seen}=\{4,6\}$}\\
Static LS & 6.961 & 3.864 & 0.377 & 2.440 & 0.1296\\
Proposed SI & 8.060 & 3.309 & 0.381 & 2.358 & 0.1266\\
Proposed SD & \textbf{8.436} & \textbf{3.125} & \textbf{0.391} & \textbf{2.263} & \textbf{0.1243}\\
\midrule
\multicolumn{6}{l}{Mic. count: $P_{\rm unseen}=\{5,7\}$}\\
Static LS & 8.012 & 3.252 & 0.398 & 2.279 & 0.1211\\
Proposed SI & 9.351 & 2.622 & 0.402 & 2.161 & 0.1180\\
Proposed SD & \textbf{9.759} & \textbf{2.445} & \textbf{0.414} & \textbf{2.042} & \textbf{0.1148}\\
\midrule
\multicolumn{6}{l}{Source count: $K_{\rm seen}=\{1,2\}$}\\
Static LS & 6.961 & 3.864 & 0.377 & 2.440 & 0.1296\\
Proposed SI & 8.060 & 3.309 & 0.381 & 2.358 & 0.1266\\
Proposed SD & \textbf{8.436} & \textbf{3.125} & \textbf{0.391} & \textbf{2.263} & \textbf{0.1243}\\
\midrule
\multicolumn{6}{l}{Source count: $K_{\rm unseen}=\{5,7\}$}\\
Static LS & 6.513 & 3.952 & 0.340 & 2.398 & 0.1327\\
Proposed SI & 7.505 & 3.392 & 0.343 & 2.340 & 0.1300\\
Proposed SD & \textbf{7.769} & \textbf{3.242} & \textbf{0.350} & \textbf{2.256} & \textbf{0.1285}\\
\bottomrule
\end{tabular*}
\end{table}

\subsection{Microphone-count generalization}
Table~\ref{tab:generalization} shows that both SI and SD improve all five metrics over Static LS, with SD achieving the best group means.

For $P_{\rm seen}=\{4,6\}$, SD exceeds SI by 0.38 dB SI-SDR. The gain remains 0.41 dB for unseen microphone counts $P_{\rm unseen}=\{5,7\}$, where SI and SD outperform Static LS by 1.34 and 1.75 dB, respectively. These gains require no changes to network parameters or output heads. Together with the lower spectral and binaural errors, they support the use of shared microphone-wise processing beyond the channel counts used in training. Higher absolute scores for unseen counts do not imply an intrinsic advantage, as the groups use different microphone configurations.

\begin{figure}[t]
\centering
\includegraphics[width=0.75\columnwidth]{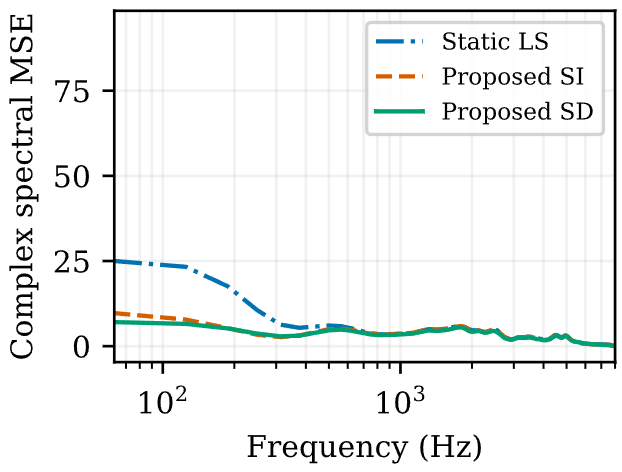}
\caption{Complex spectral MSE versus frequency on the LibriSpeech test subset with $P=6$ and $K=1$ (lower is better).}
\label{fig:frequency_mse}
\end{figure}

Fig.~\ref{fig:frequency_mse} shows that the largest absolute MSE reductions on LibriSpeech occur below approximately 200 Hz, where both neural encoders outperform Static LS and SD further improves on SI. The lower absolute errors at high frequencies may partly reflect the lower speech energy in these bands, rather than better relative reconstruction accuracy.

\subsection{Source-count generalization}
Table~\ref{tab:generalization} also compares training-seen source counts $K_{\rm seen}=\{1,2\}$ with denser mixtures at $K_{\rm unseen}=\{5,7\}$, pooling $P=4,6$ in both groups. Increasing the source count lowers SI-SDR and coherence for all three methods, indicating more difficult reconstruction of overlapping sources. Nevertheless, SI and SD retain improvements over Static LS in all five measures. SI improves SI-SDR over Static LS by 1.10 and 0.99 dB for seen and unseen source counts, respectively, while SD gains 1.47 and 1.26 dB. SI retains its relative gains more consistently as source count increases, although SD achieves higher SI-SDR in both groups.

For a fixed ATF, SI applies the same linear encoding matrix regardless of source count. It therefore preserves source superposition, although reconstruction accuracy may vary as more sources are added. SD is nonlinear because its coefficients depend on the observed signals, yet it retains an advantage over SI at unseen source counts. This supports the robustness of the ATF-conditioned framework to increased scene complexity, without isolating the contribution of signal dependence alone.

\begin{table}[t]
\centering
\caption{Cross-source-domain evaluation with $P=4,6$ and $K=5,7$, averaged equally over the four combinations.}
\label{tab:domain}
\setlength{\tabcolsep}{2pt}
\renewcommand{\arraystretch}{1.0}
\begin{tabular*}{\dimexpr\columnwidth-8pt\relax}{@{\extracolsep{\fill}}lrrr@{\hspace{0pt}}rr@{}}
\toprule
Method & SI-SDR\up & MSE\dn & Coh.\up & \shortstack{Bin.\\$\mathrm{ILD}_{\rm Err}$\dn} & \shortstack{Bin.\\$\mathrm{IC}_{\rm Err}$\dn}\\
\midrule
\multicolumn{6}{l}{\textit{Libri. (seen domain)}}\\
Static LS & 6.513 & 3.952 & 0.340 & 2.398 & 0.1327\\
Proposed SI & 7.505 & 3.392 & 0.343 & 2.340 & 0.1300\\
Proposed SD & \textbf{7.769} & \textbf{3.242} & \textbf{0.350} & \textbf{2.256} & \textbf{0.1285}\\
\midrule
\multicolumn{6}{l}{\textit{ESC-50 (unseen domain)}}\\
Static LS & 2.791 & 6.789 & 0.343 & 2.404 & 0.1336\\
Proposed SI & 3.163 & 6.533 & 0.346 & 2.349 & 0.1309\\
Proposed SD & \textbf{3.737} & \textbf{6.007} & \textbf{0.355} & \textbf{2.248} & \textbf{0.1294}\\
\midrule
\multicolumn{6}{l}{\textit{White noise (unseen domain)}}\\
Static LS & -4.268 & 12.616 & 0.317 & 2.158 & 0.1343\\
Proposed SI & -4.161 & 12.530 & 0.319 & 2.123 & 0.1324\\
Proposed SD & \textbf{-4.028} & \textbf{12.382} & \textbf{0.325} & \textbf{2.031} & \textbf{0.1320}\\
\bottomrule
\end{tabular*}
\end{table}

\subsection{Cross-source-domain generalization}
Table~\ref{tab:domain} compares LibriSpeech, ESC-50, and white noise at $P=4,6$ and $K=5,7$. ESC-50 contains diverse environmental sounds, providing a non-speech test of source-domain generalization. Although trained only on speech, both SI and SD improve all five metrics over Static LS on environmental sounds and white-noise stress tests. SD achieves the best means, gaining 1.26, 0.95, and 0.24 dB SI-SDR over Static LS in the three domains, respectively. On ESC-50, SD exceeds SI by 0.57 dB, versus 0.26 dB on LibriSpeech, and reduces MSE by 11.5\% relative to Static LS. These gains extend beyond the speech spectral statistics encountered during training. White-noise gains are smaller, and all methods have negative SI-SDR. Thus, both models still outperform Static LS on white noise, but their reconstruction quality remains limited.

\section{Conclusion}
We presented ATF-conditioned SI and SD encoders supporting variable microphone counts through shared processing and masked self-attention. Both improve on Static LS beyond the training microphone and source counts, without retraining or architectural changes. Despite training only on speech, both models retain gains on environmental sounds and white noise, although white-noise reconstruction remains challenging. SD achieves better overall reconstruction and binaural-cue metrics than SI across the evaluated conditions. These results support the framework's ability to generalize across microphone counts, source counts, and source domains. Future work will compare additional SD networks and generative methods, evaluate measured arrays and real recordings, and explore binaural-aware training and perceptual evaluation.

\clearpage
\section*{Acknowledgments}
This work was supported in part by the Science and Technology Program of Jiangsu Province under Grant BZ2024062.

\end{document}